# Microwave Sensing of Elemental Sulfur Deposition in Gas Pipelines

Mus'ab G. Magam, Hussein Attia, Member, IEEE, Khurram Karim Qureshi, Senior Member, IEEE, and Sharif I. M. Sheikh, Senior Member, IEEE

*Abstract*—A non-intrusive conformal electromagnetic-based monitoring of elemental sulfur deposition within a natural gas-carrying pipeline is presented. The deposited sulfur behaves as a superstrate layer above a sensing microstrip patch antenna that is optimally placed on the inner wall of the gas pipeline. Increasing the superstrate thickness by the sulfur deposition alters the antenna resonance behavior, which can be monitored externally. The effect of uneven or bumpy sulfur deposition is studied. Sensing antennas positioned outside a plexiglass pipeline are also investigated to observe the change in the antenna impedance matching with accumulating sulfur superstrate. Lab-based measured results agreed well with the simulated responses using commercial electromagnetic software. The proposed low-cost and easy-to-implement detection technique exhibits an accurate estimation of the deposited sulfur thickness inside the natural gas-carrying pipelines.

*Index Terms*— Elemental sulfur, Sensing Antenna, Gas Pipeline, Deposition

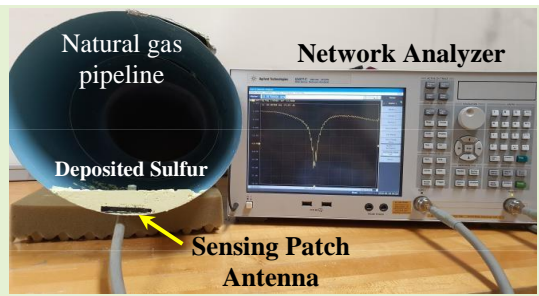

## I. Introduction

Pipelines are the primary means of transporting petroleum products such as natural gas, oil, and other vital chemicals that contribute to the world's economy. Oil and gas pipeline networks have proven to be the most cost-effective and secure method of delivering crude oil, and they meet the high demand for efficiency and dependability. In a natural gas-carrying pipeline, most solid contaminants are removed using sophisticated filtering systems in the processing plants. However, during the long-distance transmission of Hydrogen Sulfide ($H_2S$) gas produced during oil and natural gas extraction, the desublimation process produces elemental sulfur particles from its gas form [1]-[4]. Given such findings, sulfur removal techniques (pigging, chemical, and mechanical cleaning) were investigated [5]-[6]. However, such techniques incur additional costs. Therefore, it is crucial to develop a predictive method to determine elemental sulfur's early formation and deposition to avoid severe consequences.

The elemental sulfur eventually accumulates on the inner surface of the pipeline. Accumulating these unwanted elemental sulfur leads to plant shutdown by clogging the sensors, control valves, and even the pipeline. Thus, efficient monitoring of the deposited sulfur is critical for the successful delivery and metering of natural gas products [7]-[8]. Although limited studies are available on electrically monitoring the sulfur deposition rate, popular sensors used in the petrochemical industry to determine oil-carrying pipeline's flow rate and phase-fractions can be of interest. They include impedance probes, acoustic sensors, fiber-optic, and microwave sensors [9]-[11].

Compared to other methods, microwave sensors are relatively small in size, easy to install, maintain, and the reflected signal is used to obtain data about the deposition rate without interfering with the flow. Due to these advantages, the microwave measurement method has attracted many researchers [12]-[15]. A non-invasive and straightforward microwave sensor, reported in [16], used two simple antennas to identify the ratio of the three-phase (gas, brine, oil) contents within a petroleum-carrying pipeline. The first antenna is designed to detect the low-loss gas and oil ratio, while the second antenna detects the lossy brine contents of the stratified flow. According to [17], a non-invasive microwave measurement system was developed to monitor the reflected microwave signal from a gas-solid two-phase flow in a vertical pipeline for different particle properties and sizes. A microwave system with a pair of patch antenna sensors and a coaxial probe was developed in [18] to measure oil, gas, and water fraction in a pipeline. The microstrip patch antennas in transmission mode and coaxial probe in reflection mode are used to estimate the three-phase flow. This technique uses complicated empirical and physics-based models. A circular waveguide with a resonant iris structure attached to the aperture was designed in [19] to detect and monitor solid contaminants in gas pipelines. The concentration of contaminants is determined by analyzing

Manuscript received xxx; accepted xxx. Date of publication xxx; date of current version xxx. (Corresponding author: Hussein Attia, hattia@kfupm.edu.sa)

This work was supported by the Center for Communication Systems and Sensing at King Fahd University of Petroleum & Minerals (KFUPM) through project No. INCS2105.

M. G. Magam, H. Attia, K. K. Qureshi, and S. I. M. Sheikh are with the Center for Communication Systems and Sensing, and the Electrical Engineering Dept at KFUPM, Dhahran, 31261, Saudi Arabia.

the reflection coefficient of the transmitting and receiving probes at a specific resonant frequency. In [20], an invasive microwave method was used to accurately monitor sand's onset, and deposition rate within a petroleum pipeline, where properly integrating the probe with minimum leakage remains a challenge. A microwave system was designed in [21]-[23], where microwave transmission and reflection responses are analyzed to detect the presence of black power contaminants within the natural gas carrying pipeline. This process involved complicated measurements and analysis processes since both the magnitude and phase responses were recorded and analyzed.

In this paper, multiple conformal antenna resonators are optimally integrated internally and externally to detect sulfur deposition within a long-distance natural gas-carrying pipeline. The elemental sulfur deposited on the microstrip antenna's surface is considered a superstrate, which affects the antenna resonance characteristics. Thus, with increasing superstrate thickness, the antennas mounted on the inner wall of a metallic pipeline demonstrated decreasing resonant frequencies. In comparison, the antennas mounted on the outer wall of a plexiglass pipeline demonstrated an increased reflected power level (~ $|S_{11}|$). Although more accurate monitoring is achieved when integrating antennas within the inner wall of the pipeline, antennas placed outside the natural gas pipeline demonstrated an advantage of single-frequency operation. This is important to achieve the design trade-offs between measurement accuracy and difficulty in design and fabrication.

## II. DESIGN AND ANALYSIS

Conformal microstrip antenna resonators are popular in microwave sensors due to their low profile, lightweight, easy fabrication and integration [24]. A dielectric cover over the radiating microstrip antenna or array is generally used for protection from environmental hazards in many applications. These covers are often realized with single or multilayer superstrates, which change the resonance behavior of the original antenna. In [25]-[26], the effect of superstrate loading on the antenna's resonant frequency, gain, directivity, and polarization was investigated. It is demonstrated that different types of antenna loading have different influences on the resonance behavior. Direct and indirect superstrate loading mainly affects the effective dielectric constant and the co- and cross-polarization ratio of the antenna, introducing the changes in the resonance frequency and the input impedance [25]. Nevertheless, to accurately predict the loading effect of the superstrate on the antenna resonance frequency, the existing antenna model needs to include the superstrate's dielectric parameters, as depicted in the following equations [27]

$$\frac{\Delta f_r}{f_r} = \frac{1}{2} \frac{\Delta \varepsilon_e / \varepsilon_{eo}}{1 + \frac{1}{2} \Delta \varepsilon_e / \varepsilon_{eo}} \quad (1)$$

$$\varepsilon_e = \varepsilon_{eo} + \Delta \varepsilon_e \quad (2)$$

where $\varepsilon_e$ is the effective dielectric constant with the superstrate, $\varepsilon_{eo}$ is the effective dielectric constant without superstrate, $\Delta \varepsilon_e$ represents the change in dielectric constant due to superstrate, and $\frac{\Delta f_r}{f_r}$ is the fractional resonant frequency due to the superstrate. Numerical results in [27] show that the effective dielectric constant is highly affected by a superstrate material with a high dielectric constant. Furthermore, the change is more noticeable for small ratios between the width to height (W/h) of the substrate material. This knowledge is essential in designing and optimizing the antenna resonator to monitor the sulfur superstrate thickness inside a gas-carrying pipeline.

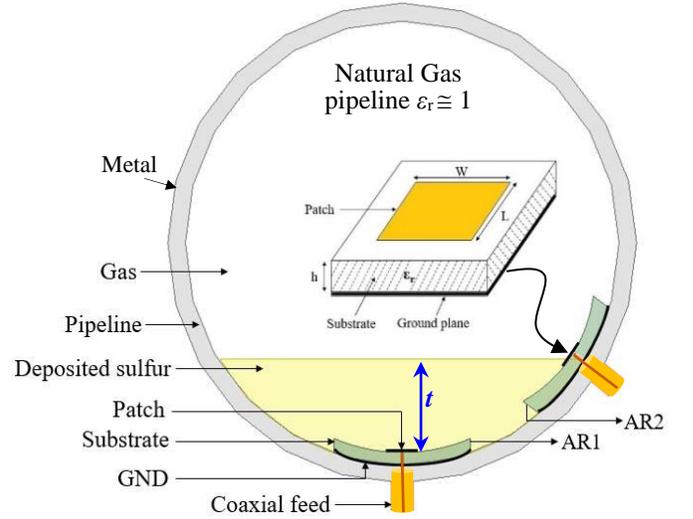

Fig. 1. Schematic diagram of the natural gas-carrying pipeline with two internally integrated antenna resonators (AR1 and AR2) and deposited elemental sulfur of maximum thickness (t).

In our proposed design and as depicted in Fig. 1, the employed sensing element is a simple and low-cost microstrip patch antenna with a rectangular shape and operating at 10 GHz. The frequency of operation is chosen to have a reasonable-size and easy-to-fabricate antenna.

The radiating patch element is printed on a low-loss Rogers RT/Duroid 5880 substrate. The substrate material has a relative permittivity of 2.2, a loss tangent of 0.0009, and a thickness of 0.508 mm. The commercial full-wave electromagnetic software, 3D High-Frequency Simulation Software (HFSS), is used to optimize the antenna performance before the experimental verification. A coaxial-line feeder with a characteristic impedance of 50-ohm is used to feed the sensing microstrip antenna. As seen in Fig. 1, the inner conductor of the coaxial-feeder is extended through the metallic pipeline and substrate and connected to the antenna's radiating patch, while the outer conductor is soldered to the antenna ground plane. The key benefits of this feeding mechanism are its ease of fabrication and assembly in this specific application, good impedance matching with the radiating element (i.e., less reflected power), and minimal spurious radiation.

The sensing microstrip antenna with a rectangular shape is designed to operate at 10 GHz and has dimensions of W= 9.7 mm and L= 11.69 mm (see inset of Fig. 1). The antenna is optimally integrated within the inner wall of a metallic natural gas-carrying pipeline with a diameter of 140 mm, as shown in Fig. 1. Note that the deposited sulfur superstrate layer is assumed to have a maximum thickness $t$ and $\varepsilon_r$ = 3.5 [28]. The surrounding natural gas with the majority of its contents (80-95%) as Methane (CH4) has $\varepsilon_r \cong 1$ [29] as shown in Fig. 1. As the thickness of the deposited sulfur increases, multiple antenna resonators (AR1 and AR2) are used to detect the deposited sulfur's thickness accurately. The natural gas contents of the pipeline have little effect on the antenna resonance due to their separation from the antenna surface, and, also, it acts as free-space with $\varepsilon_r \cong 1$. Since the top surface of the deposited sulfur may not be flat in reality, antenna responses for superstrates with uniform (flat) and non-uniform (wavy) deposited sulfur are analyzed in the next sections.

### A. Uniform deposition of elemental sulfur

In this sub-section, the change of the sensing element's resonance response (magnitude and phase) is monitored for different thicknesses of the elemental sulfur superstrate. The frequency responses related to the thickness of the deposited sulfur for non-invasive and invasive single and multiple antenna sensors are investigated.

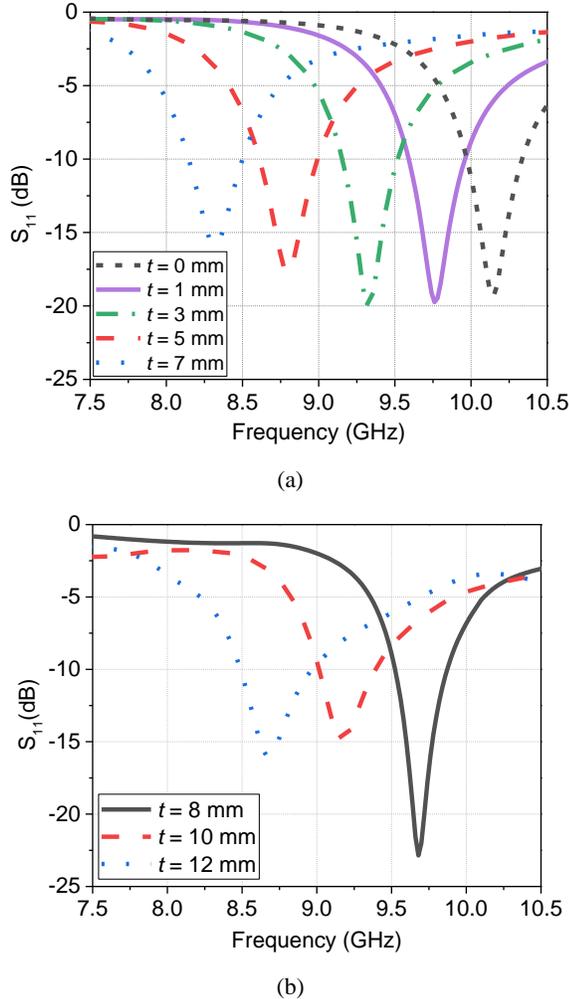

Fig. 2. Effect of different thicknesses (t) of the deposited elemental sulfur on the reflection coefficient of (a) Antenna resonator (AR1), and (b) Antenna resonator (AR2) used for thicker layers of sulfur.

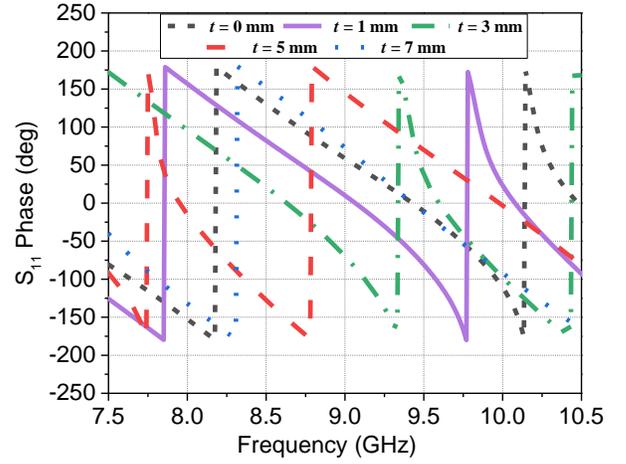

Fig. 3. $S_{11}$ phase response produced by the antenna resonator AR1 for different thicknesses (t) of the deposited elemental sulfur superstrate.

Figure 2(a) displays the simulated frequency response of an invasive antenna AR1 (seen in Fig. 1) for thinner layers of the deposited sulfur superstrate. Note that increasing sulfur superstrate thickness from onset to 7 mm resulted in decreasing the antenna resonance frequency from 10.15 GHz to 8.3 GHz (22.3%). Also, a resonance frequency shift of about 0.5 GHz for each 2 mm thickness increase of deposited sulfur is noticed. A second antenna resonator (AR2), shown in Fig. 1, is used to detect thicker layers of the deposited sulfur due to its elevated position with respect to AR1. Figure 2(b) plots the reflection coefficient of AR2 for deposited sulfur thickness ranging from 8 to 12 mm. A decrease in the antenna resonance frequency from 9.7 GHz to 8.7 GHz is observed.

Additional antennas can be placed across the remaining perimeter of the inner wall to detect a more advanced stage of sulfur deposition within the pipeline. The related phase response of the antenna is plotted in Fig. 3, which also reflects the changing resonant behavior with different layer thicknesses of sulfur superstrate. Although this study considers that unwanted solid contaminants other than sulfur (such as sand and black-powder) are removed by filtering, analyzing the complete reflection responses of Figs 2(a) and 3 together can allow identifying the presence of other contaminants.

Although accurate, the invasive nature of this monitoring technique involves drilling antenna access holes through the metal pipeline. This may cause leakage problems due to the pressurized gas flow in the long run. To remedy this situation, a non-invasive technique is also investigated here, where the antenna is placed outside a plexiglass (non-metal) pipeline. Figure 4 shows the schematic diagram of this setup, where it is clear that this setup can be easily transported to different locations. Since the thickness of the plexiglass pipeline affects the detection sensitivity, unlike the invasive technique, this method uses a simpler single-frequency operation.

Figures 5(a) and (b) show the effect of different thicknesses (t) of the deposited elemental sulfur on the reflection response ($S_{11}$) of the externally mounted antenna on the outer wall of the plexiglass pipeline. It is clear from Fig. 5(a) that the higher the thickness of the deposited sulfur, the higher the reflected power, which is expected and agrees with intuition. Also, more phase variation of ($S_{11}$) around the resonance frequency of 10.1 GHz is observed for less thickness of the deposited sulfur, as depicted in Fig. 5(b). Since accurate detection of power level is challenging in practice, combined analysis of $S_{11}$ magnitude and phase measurements can be used to increase the reliability of the measurements.

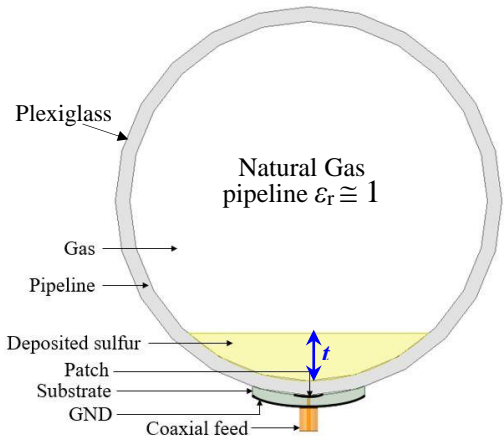

Fig. 4. Schematic diagram of the natural gas pipeline with an external antenna mounted on the outer wall of the plexiglass pipeline.

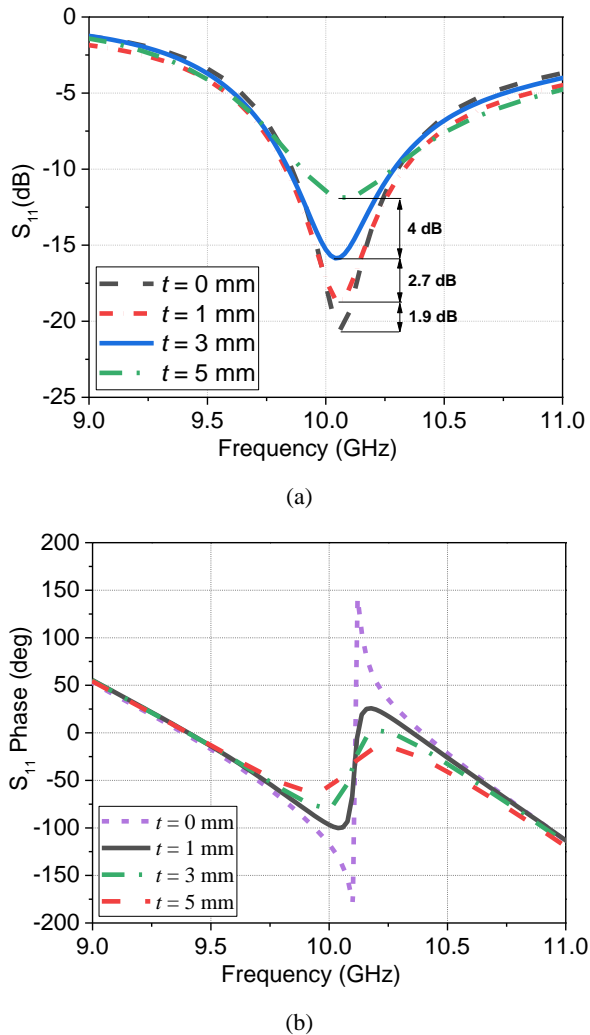

Fig. 5. Effect of different thicknesses ($t$) of the deposited elemental sulfur on the reflection response ($S_{11}$) of the externally mounted antenna on the outer wall of a plexiglass pipeline (a) Magnitude, and (b) Phase response.

### B. Non-uniform deposition of elemental sulfur

This section presents the antenna resonator (AR1) response for the non-uniform (wavy) top edge of the sulfur superstrate. Figure 6(a) shows a schematic diagram of the natural gas pipeline with an internally integrated microstrip resonator antenna and a non-uniformly deposited elemental sulfur. The reflection response of the antenna resonator for different thicknesses ($t$) of the non-uniform superstrate is shown in Fig. 6(b). Note that increasing sulfur deposition from 1 mm to 5 mm changes the resonant frequency from 9.6 GHz to 8.7 GHz, which is similar to the uniform-case response shown in Fig. 2(a). Also, Fig. 6(b) shows that the higher the thickness of the deposited sulfur, the higher the reflected power. Note that similar responses are noted for other non-uniform top edge configurations (i.e., other wavy shapes of the top edge of the sulfur). It is worth mentioning that for non-uniform deposition, the reflection response saturates faster than the uniform case, and additional antennas are needed to monitor thicker depositions of the sulfur.

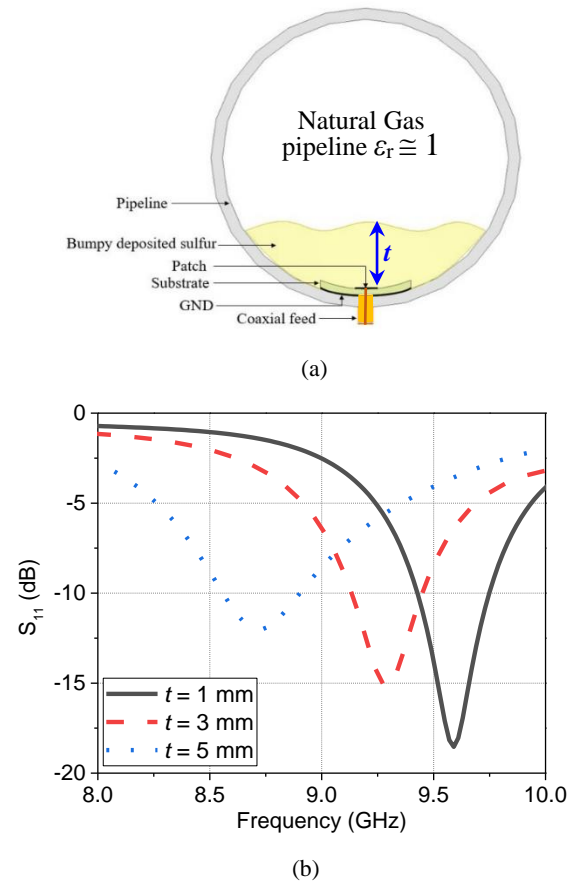

Fig. 6. (a) Schematic diagram of a wavy-like deposited sulfur layer within the natural gas pipeline (b) Reflection coefficient response for different thicknesses ($t$) of the sulfur layer.

## III. RESULTS AND DISCUSSION

To verify the simulated results, the employed microstrip antenna sensor is fabricated and integrated into the inner wall of the gas pipeline. Figure 7(a) shows the top view of the fabricated microstrip antenna resonator, and Fig. 7(b) shows the side-view of the antenna with accumulated sulfur superstrate. The experimental setup with the sensing antenna integrated into the inner wall of the lab-based pipeline segment is shown in Fig. 8. To avoid $H_2S$ gas-related poisoning, the sulfur powder was manually deposited in the pipe to represent the accumulated elemental sulfur superstrate above the sensing microstrip antenna. A uniform deposition scenario was considered in the experiment for simplicity. A calibrated vector network analyzer (Agilent E5071C) that supports up to 20 GHz measurements, shown in Fig. 8, is used to record the changing reflection response ($S_{11}$) with increasing the superstrate thickness from 1-5 mm. A superimposed plot comparing the simulated reflection coefficient with the experimental results is shown in Fig. 9. Note that for increasing the deposition thickness from 1-5 mm, the experimental $S_{11}$ responses agreed reasonably with the simulated values. A resonance frequency shift of about 0.5 GHz for each 2 mm thickness increase of deposited sulfur is noticed. The slight discrepancy between the simulated and measured results can be attributed to the unavoidable human errors in manually measuring the superstrate height or due to the ideal approximations of the numerical full-wave simulator HFSS.

## IV. CONCLUSION

In long-distance natural gas-carrying pipelines, elemental sulfur deposition is of paramount concern as other solid contaminants are removed by filtering. Detecting the onset of uniform or non-uniform sulfur deposition is essential to avoid clogging the sensors and control valves that lead to extensive downtime of plant operation. A simple microwave approach to sensing the presence of elemental sulfur is proposed using direct and indirect loading of optimally embedded microstrip antenna resonators. Accumulated elemental sulfur behaves as a superstrate and affects the antenna resonant properties. For antennas integrated into the inner wall of the gas pipeline, the resonant frequencies are observed to decrease with increasing superstrate thickness. For non-invasive antennas placed on the outer wall of the gas pipeline, the decreasing magnitude of a single-frequency reflection coefficient indicates more reflected power with increasing the deposited sulfur thickness. The simulated and experimental $S_{11}$ responses for different thicknesses of the deposited elemental sulfur agreed well for uniform deposition. Minor disagreements between these results are mainly due to human error in reading sulfur superstrate thickness. Multiple antenna sensors can be used to monitor the level of the deposited sulfur across the perimeter of the natural gas pipeline.

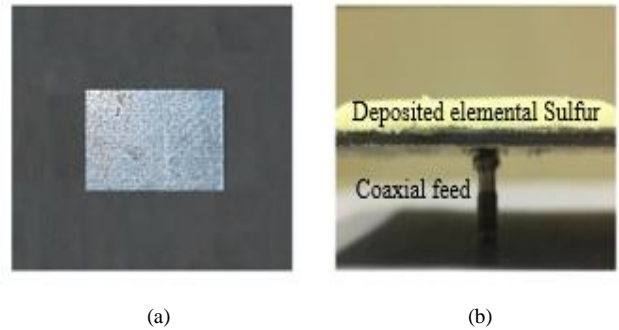

Fig. 7. (a) Top-view of the fabricated 10 GHz coaxial-line-fed microstrip antenna (b) Side-view of the antenna covered with a deposited elemental sulfur superstrate.

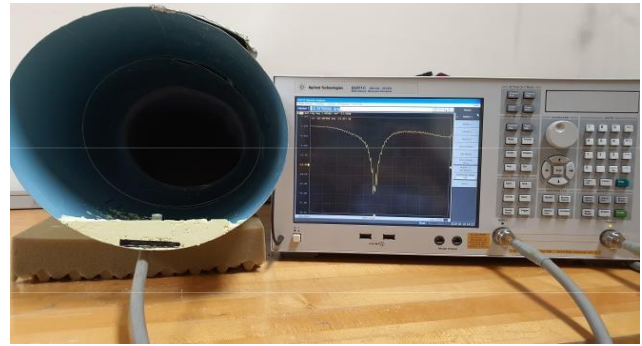

Fig. 8. Measurement setup of the reflection coefficient ($S_{11}$) of the sensing microstrip antenna covered with uniform sulfur superstrate, observed using a calibrated Vector Network Analyzer (Agilent E5071C).

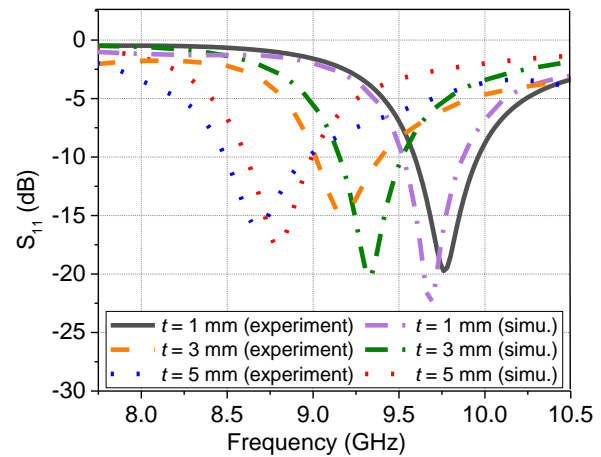

Fig. 9. Superimposed measured and simulated reflection response due to different thicknesses (t) of the uniform elemental sulfur deposition.

## ACKNOWLEDGMENT

The authors acknowledge the support from the Center for Communication Systems and Sensing at King Fahd University of Petroleum & Minerals (KFUPM) through project No. INCS2105.

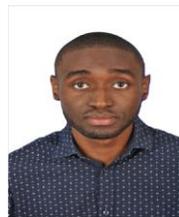

**M.G. Magam** received the B.Sc. degree in electronics and communication engineering from Arab Academy for Science, Technology and Maritime Transport, Cairo, Egypt, in 2015. He recently completed his master's degree in electrical engineering at King Fahd University of Petroleum and Minerals, Dhahran, Saudi Arabia. His research interest includes the design and modeling of microwave devices, microwave and optical detection of sulfur in oil and gas pipelines, 4G/5G Antennas, and optical fiber sensors.

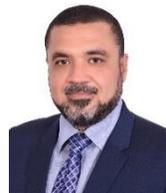

**Hussein Attia** (Member, IEEE) received a Ph.D. degree in electrical and computer engineering from the University of Waterloo, Waterloo, ON, Canada, in 2011. He worked as a Research Engineer with the Coding and Signal Transmission Laboratory, University of Waterloo, from March 2011 till July 2013. He was granted a Postdoctoral Fellowship at Concordia University, Montreal, QC, Canada, from August 2014 to July 2015. Also, he was a Visiting Scholar with University de Quebec (INRS) from August 2015 to December 2015 and from June 2017 to August 2017. He is currently an Associate Professor with the King Fahd University of Petroleum and Minerals (KFUPM). He published 80+ journal and conference papers. His research interests include millimeter-wave and wide-band antennas, EM sensors for biomedical applications, analytical techniques for electromagnetic modeling, and engineered magnetic metamaterials. During his Ph.D. program, he received the University of Waterloo Graduate Scholarship for excellence in research and coursework in 2009. He was a finalist in the Student Paper Competition of the 2011 IEEE AP-S International Symposium on Antennas and Propagation. He was ranked first among all B.Sc. students of electronics and communication engineering, Zagazig University, Egypt, in 1999.

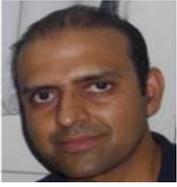

**Khurram Karim Qureshi** (SM'12) received a BSc degree with honors in Electrical Engineering from the University of Engineering and Technology (UET), Lahore, Pakistan, and a Ph.D. also in Electrical Engineering from the Hong Kong Polytechnic University in 2006. He is currently an Associate Professor with the Electrical Engineering Department of King Fahd University of Petroleum and Minerals (KFUPM). His research interests include optical communications, optical signal processing, lasers, sensors, and miniaturized antennas. He is a senior member of IEEE, USA, and has published more than 80 journal and conference papers and five US patents issued to his credit.

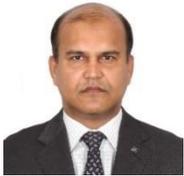

**Sharif Iqbal** completed his MSc and Ph.D. degrees at the University of Manchester (UMIST), UK. He is currently employed as an Associate Professor at the Department of Electrical Engineering, KFUPM, Saudi Arabia. He published 100+ scholarly works in refereed journals, conferences, patents, etc. He is a Fellow of IEE and a Senior Member of IEEE.